\documentclass[aps,prl,twocolumn,groupedaddress]{revtex4-2}

\usepackage{ulem}
\usepackage{amsmath}
\usepackage{bbold}
\usepackage{bm}
\usepackage{graphicx}
\usepackage{hyperref}
\usepackage{times}
\hypersetup{colorlinks=true,citecolor=magenta,linkcolor=magenta,urlcolor=magenta}

\begin{document}


\title{State-dependent error bound for digital quantum simulation of driven systems}


\author{Takuya Hatomura}
\email[]{takuya.hatomura.ub@hco.ntt.co.jp}
\affiliation{NTT Basic Research Laboratories \& NTT Research Center for Theoretical Quantum Physics, NTT Corporation, Kanagawa 243-0198, Japan}


\date{\today}

\begin{abstract}
Digital quantum simulation is a promising application of quantum computers, where quantum dynamics is simulated by using quantum gate operations. 
Many techniques for decomposing a time-evolution operator of quantum dynamics into simulatable quantum gate operations have been proposed, while these methods cause some errors. 
To evaluate these errors, we derive a lower bound for overlap between true dynamics and digital simulated dynamics at the final time. 
Our result enables us to guarantee how obtained digital simulated dynamics is close to unknown true dynamics. 
We also extend our formalism to error evaluation of digital quantum simulation on noisy quantum computers. 
\end{abstract}


\maketitle


\textit{Introduction.---}
Simulation of quantum dynamics is a basic approach for deep understanding of quantum nature. 
However, its implementation on classical computers is limited since it generally requires exponentially large degrees of freedom against system size. 
Quantum computers can simultaneously handle exponentially large number of orthogonal states as quantum superpositions. 
This is why simulation of quantum dynamics on quantum computers was expected at the dawn of quantum information science~\cite{Feynman1982}. 
After that, it was shown that this expectation is correct, i.e., quantum computers can efficiently simulate quantum dynamics in a discrete way~\cite{Lloyd1996}.

In digital quantum simulation, time-evolution operators of quantum dynamics are divided into short time slices and each slice is further decomposed into a series of simulatable quantum gate operations~\cite{Lloyd1996,Wiebe2010,Poulin2011,Somma2016,Hadfield2018,Childs2019,Childs2021}. 
Decomposition is exact for infinitesimal time slices, but in practice such decomposition is impossible and finite decomposition causes errors~\cite{Suzuki1976,Suzuki1985,Suzuki1991,Huyghebaert1990}. 
The amount of errors during each time slice has been evaluated in terms of time-evolution operators for true dynamics and digital simulated dynamics~\cite{Lloyd1996,Wiebe2010,Poulin2011,Somma2016,Hadfield2018,Childs2019,Childs2021}. 
However, this error evaluation may overestimate contributions which are irrelevant to dynamics.

To avoid this overestimation, we should directly consider difference between true dynamics and digital simulated dynamics, while information of true dynamics cannot be used since finding it is nothing but the purpose of quantum simulation. 
Recently, it was found that lower bounds for overlap between two different dynamics can be calculated by using one of these two dynamics~\cite{Suzuki2020,Hatomura2021,Funo2021,Hatomura2021a,Takahashi2021}. 
These results enable us to evaluate overlap between unknown true dynamics and known approximate dynamics by using only known approximate dynamics (see, Sec. II. A of Ref.~\cite{Hatomura2021a}) although their formalism cannot straightforwardly be applied to the present problem.

In this paper, we introduce distance between true dynamics and its digital simulated dynamics at the final time. 
Then, we derive its upper bound which is equivalent to a lower bound for overlap between these two dynamics. 
We point out that our result gives a more straightforward and tighter bound for error evaluation of digital quantum simulation than conventional approaches. 
We also extend our formalism to error evaluation of noisy digital quantum simulation, i.e., digital quantum simulation on noisy quantum computers.

\textit{Digital quantum simulation.---}
We consider obtaining a target state $|\Psi(T)\rangle$ generated by a time-evolution operator
\begin{equation}
\begin{aligned}
\hat{U}(T,0)&=\mathcal{T}\exp\left(-\frac{i}{\hbar}\int_{0}^Tdt\hat{\mathcal{H}}(t)\right)\\
&=\lim_{M\to\infty}\prod_{m=M}^1\exp\left(-\frac{i}{\hbar}\frac{T}{M}\hat{\mathcal{H}}(mT/M)\right)\\
&\equiv\lim_{M\to\infty}\prod_{m=M}^1\hat{\mathcal{U}}(mT/M,(m-1)T/M),
\end{aligned}
\end{equation}
i.e., $|\Psi(T)\rangle=\hat{U}(T,0)|0\rangle$, where $T$ is total operation time, $\hat{\mathcal{H}}(t)$ is a time-dependent Hamiltonian, $\mathcal{T}$ is the time-ordering operator, and $|0\rangle$ is an initial state. 
In digital quantum simulation, we stroboscopically simulate this dynamics by
\begin{equation}
\begin{aligned}
|\Phi(nT/M)\rangle=\prod_{m=n}^1\hat{\mathcal{U}}(mT/M,(m-1)T/M)|0\rangle,&\\
n=1,2,\dots,M,&
\end{aligned}
\end{equation}
with finite, but large $M$ compared with $T$, where each sub-factor of the time-evolution operator $\hat{\mathcal{U}}(mT/M,(m-1)T/M)$ can be decomposed into simulatable quantum gate operations since it just puts time forward by a small time interval $T/M$.

\textit{Error bound.---}
In conventional approaches, the amount of errors is evaluated by
\begin{equation}
\|\hat{U}(nT/M,(n-1)T/M)-\hat{\mathcal{U}}(nT/M,(n-1)T/M)\|,
\label{Eq.error.norm}
\end{equation}
where $\|\cdot\|$ is a certain norm, e.g., the spectral norm. 
However, it may overestimate errors which are irrelevant to simulated dynamics since the norm gives the worst  case deviation among all possible states. 
Note that only recently this point was taken into account~\cite{Somma2016}, but its formalism still requires information of true dynamics.

We rather focus on distance between the digital simulated dynamics $|\Phi(T)\rangle$ and the true dynamics $|\Psi(T)\rangle$, i.e., the Fubini-Study angle
\begin{equation}
\mathcal{L}(|\Phi(T)\rangle,|\Psi(T)\rangle)=\arccos|\langle\Phi(T)|\Psi(T)\rangle|,
\label{Eq.dist}
\end{equation}
which is the statistical distance between two quantum states~\cite{Wootters1981}. 
From the definition, it can be rewritten as $\mathcal{L}(|\Phi(T)\rangle,|\Psi(T)\rangle)=\mathcal{L}(|\chi(T)\rangle,|0\rangle)$, where $|\chi(nT/M)\rangle=[\hat{U}(nT/M,0)]^\dag|\Phi(nT/M)\rangle$. 
Since $|\chi(0)\rangle=|0\rangle$, we find an inequality $\mathcal{L}(|\chi(T)\rangle,|0\rangle)\le\sum_{n=1}^M\mathcal{L}(|\chi(nT/M)\rangle,|\chi((n-1)T/M)\rangle)$ by using the triangle inequality of the distance. 
Finally, we obtain an upper bound
\begin{equation}
\mathcal{L}(|\Phi(T)\rangle,|\Psi(T)\rangle)\le\sum_{n=1}^M\mathcal{L}_n,
\end{equation}
where
\begin{equation}
\begin{aligned}
\mathcal{L}_n=&\arccos|\langle\Phi(nT/M)|\hat{U}(nT/M,(n-1)T/M)\\
&\times[\hat{\mathcal{U}}(nT/M,(n-1)T/M)]^\dag|\Phi(nT/M)\rangle|.
\end{aligned}
\end{equation}
This upper bound for the distance (\ref{Eq.dist}) is equivalent to a lower bound for overlap between the digital simulated dynamics and the true dynamics. 
Indeed, we can rewrite it as
\begin{equation}
|\langle\Phi(T)|\Psi(T)\rangle|\ge\cos\left(\sum_{n=1}^M\mathcal{L}_n\right),
\label{Eq.lower.overlap}
\end{equation}
for $\sum_n\mathcal{L}_n\le\pi/2$ (it results in a trivial bound $|\langle\Phi(T)|\Psi(T)\rangle|\ge0$ for $\sum_n\mathcal{L}_n>\pi/2$).

The advantage of our result is that the deviation of the digital simulated dynamics from the true dynamics at the final time~(\ref{Eq.dist}) is straightforwardly evaluated instead of the difference in the time-evolution operators for each time slice~(\ref{Eq.error.norm}). 
Moreover, our result is tighter than conventional approaches. 
In conventional approaches, we discuss how Eq.~(\ref{Eq.error.norm}) is close to $0$. 
In our case, we discuss how $\cos\mathcal{L}_n$ is close to $1$. 
Since $\|\cdot\|=\max_{|\psi\rangle}\|\cdot|\psi\rangle\|\ge\|\cdot|\Phi\rangle\|$ and $\mathrm{Re}(\cdot)\le|\cdot|$, we find
\begin{equation}
[\text{Eq.~(\ref{Eq.error.norm})}]\ge\sqrt{2-2\cos\mathcal{L}_n}. 
\end{equation}
Note that this inequality approximately gives $[\text{Eq.~(\ref{Eq.error.norm})}]\gtrsim\mathcal{L}_n$ for small $\mathcal{L}_n\ll1$. 
In addition, since $0\le\sqrt{2}(1-\cos\mathcal{L}_n)\le\sqrt{2-2\cos\mathcal{L}_n}$, we obtain
\begin{equation}
[\text{Eq.~(\ref{Eq.error.norm})}]\ge1-\cos\mathcal{L}_n\ge0.
\end{equation}
Namely, $1-\cos\mathcal{L}_n$ is closer to $0$ than Eq.~(\ref{Eq.error.norm}). 
Note that later we also confirm tightness of our bound by numerical simulation.

Now we calculate each distance $\mathcal{L}_n$. 
For simplicity, we use the first-order Suzuki-Trotter decomposition, but it can easily be sophisticated by adopting higher-order decomposition or other expansion techniques (see the state-of-the-art technique for Eq.~(\ref{Eq.error.norm}) in Ref.~\cite{Childs2021} and references therein). 
Suppose that the Hamiltonian is given by
\begin{equation}
\hat{\mathcal{H}}(t)=\sum_{k=1}^K\hat{H}_k(t), 
\end{equation}
where dynamics under each sub-Hamiltonian $\hat{H}_k(t)$ is simulatable, and the Hamiltonian $\hat{\mathcal{H}}(t)$ can be regarded as a time-independent Hamiltonian within each slice, i.e., $\hat{\mathcal{H}}(mt/M)\approx\hat{\mathcal{H}}((m-1)t/M)$. 
Since the time interval $T/M$ is small, we find
\begin{equation}
\begin{aligned}
&\mathcal{L}_n\approx\frac{T^2}{2\hbar^2M^2}\Bigg|\langle\Phi(nT/M)|\hat{A}(nT/M)|\Phi(nT/M)\rangle\Bigg|, \\
&\hat{A}(nT/M)=\sum_{\substack{k,l=1 \\ (k>l)}}^K[\hat{H}_k(nT/M),\hat{H}_l(nT/M)],
\end{aligned}
\label{Eq.dist.app}
\end{equation}
up to the second order of $T/M$. 
This is of course consistent with the previous result using the first-order Suzuki-Trotter decomposition~\cite{Poulin2011}, but it is improved by replacing the norm with the absolute value of the expectation value. 
As mentioned above, we can also improve other results in a similar way. 
Notably, without information of the true dynamics, Eqs.~(\ref{Eq.lower.overlap}) and (\ref{Eq.dist.app}) provide an approximate lower bound for the overlap between these two dynamics $|\langle\Phi(T)|\Psi(T)\rangle|$.

\textit{Example.---}
Now we demonstrate how our result improves the previous result. 
We consider digital quantum simulation of quantum annealing in the transverse-field Ising chain. 
The Hamiltonian is given by
\begin{equation}
\hat{\mathcal{H}}(t)=-\frac{t}{T}\sum_{i=1}^L\hat{Z}_i\hat{Z}_{i+1}-\left(1-\frac{t}{T}\right)\sum_{i=1}^L\hat{X}_i,
\end{equation}
where $\{\hat{X}_i,\hat{Y}_i,\hat{Z}_i\}_{i=1}^L$ is a set of the Pauli matrices for $L$ qubits. 
To adopt the conventional notation of quantum annealing, we set $\hbar=1$. 
Note that we adopt the periodic boundary condition, $\hat{W}_{L+1}=\hat{W}_1$ ($W=X,Y,Z$), and a dimensionless expression, i.e., we assume that the interaction term and the transverse field term have a same energy scale and it is omitted. 
In digital quantum simulation, the time-evolution operator for each time slice is decomposed into the interaction term and the transverse field term. 
Then, we obtain
\begin{equation}
\hat{A}(nT/M)=\pm2i\frac{n}{M}\left(1-\frac{n}{M}\right)\sum_{i=1}^L(\hat{Y}_i\hat{Z}_{i+1}+\hat{Z}_i\hat{Y}_{i+1}), 
\end{equation}
where the sign depends on the order of decomposition, but it does not affect the final conclusion.

\begin{figure}
\includegraphics[width=8cm]{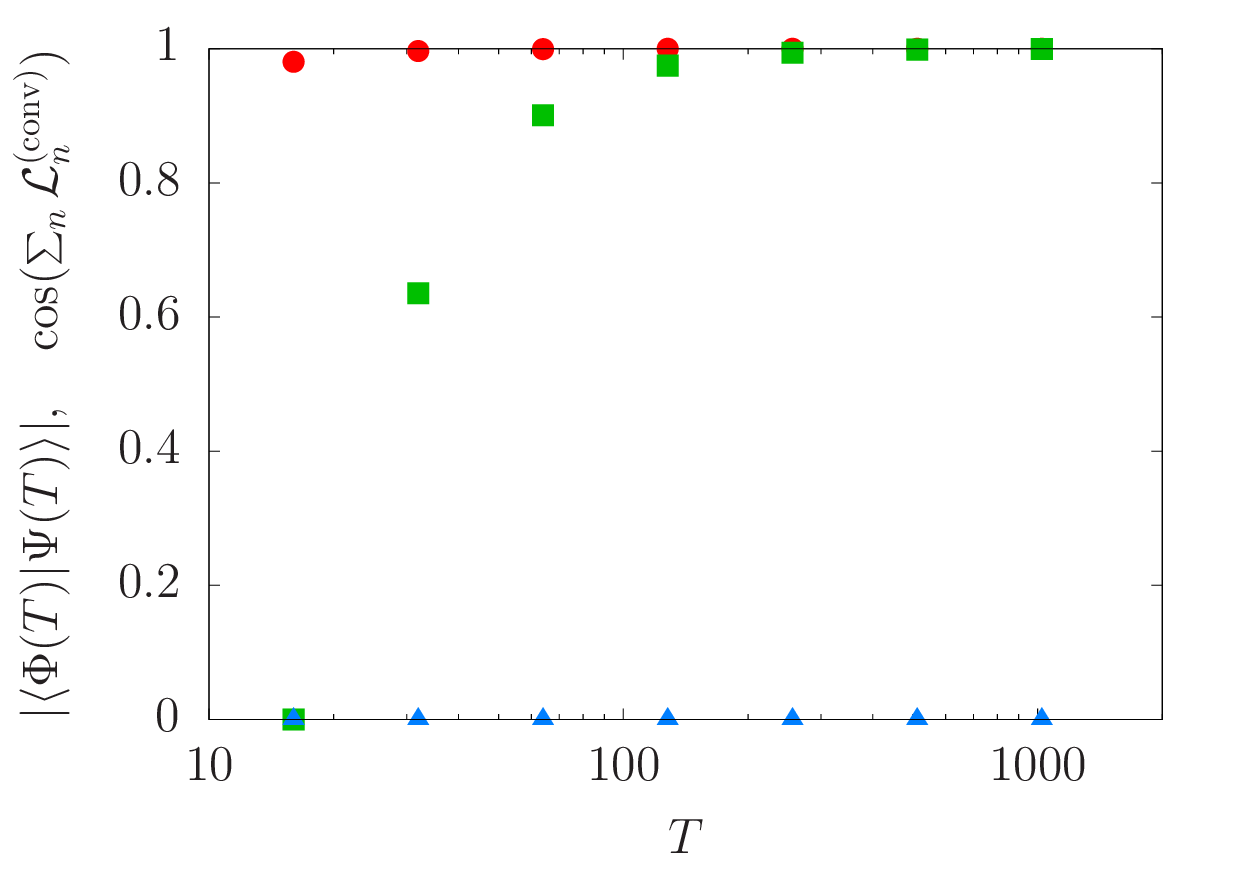}
\caption{\label{Fig.bd}Overlap between the digital simulated dynamics and the true dynamics with its lower bounds against the annealing time $T$. Each symbol represents (red circles) the overlap $|\langle\Phi(T)|\Psi(T)\rangle|$, (green squares) our bound $\cos(\sum_n\mathcal{L}_n)$, and (blue triangles) the conventional bound $\cos(\sum_n\mathcal{L}_n^{\mathrm{conv}})$, respectively. Here, $M=T^2$. }
\end{figure}

Now we numerically compare our result (\ref{Eq.dist.app}) with the conventional one
\begin{equation}
\mathcal{L}_n^\mathrm{conv}=\frac{T^2}{2M^2}\|\hat{A}(nT/M)\|,
\end{equation}
where $\|\cdot\|$ is the spectral norm. 
In numerical simulation, we set $L=100$, and then there are two variables, the annealing time $T$ and the number of the time slices $M$. 
First, we change the annealing time $T$. 
The total amount of the errors is roughly scaled as $T^2/M$ since the amount of the errors for each slice is roughly scaled as $T^2/M^2$ and there are $M$ time slices. 
Here we set $M=T^2$, for which digital quantum simulation using the first-order Suzuki-Trotter decomposition may be broken down. 
In Fig.~\ref{Fig.bd}, the overlap between the digital simulated dynamics and the true dynamics for various annealing time $T$ is plotted with the lower bounds. 
Our bound becomes tight for large annealing time, while the conventional bound is useless.

\begin{figure}
\includegraphics[width=8cm]{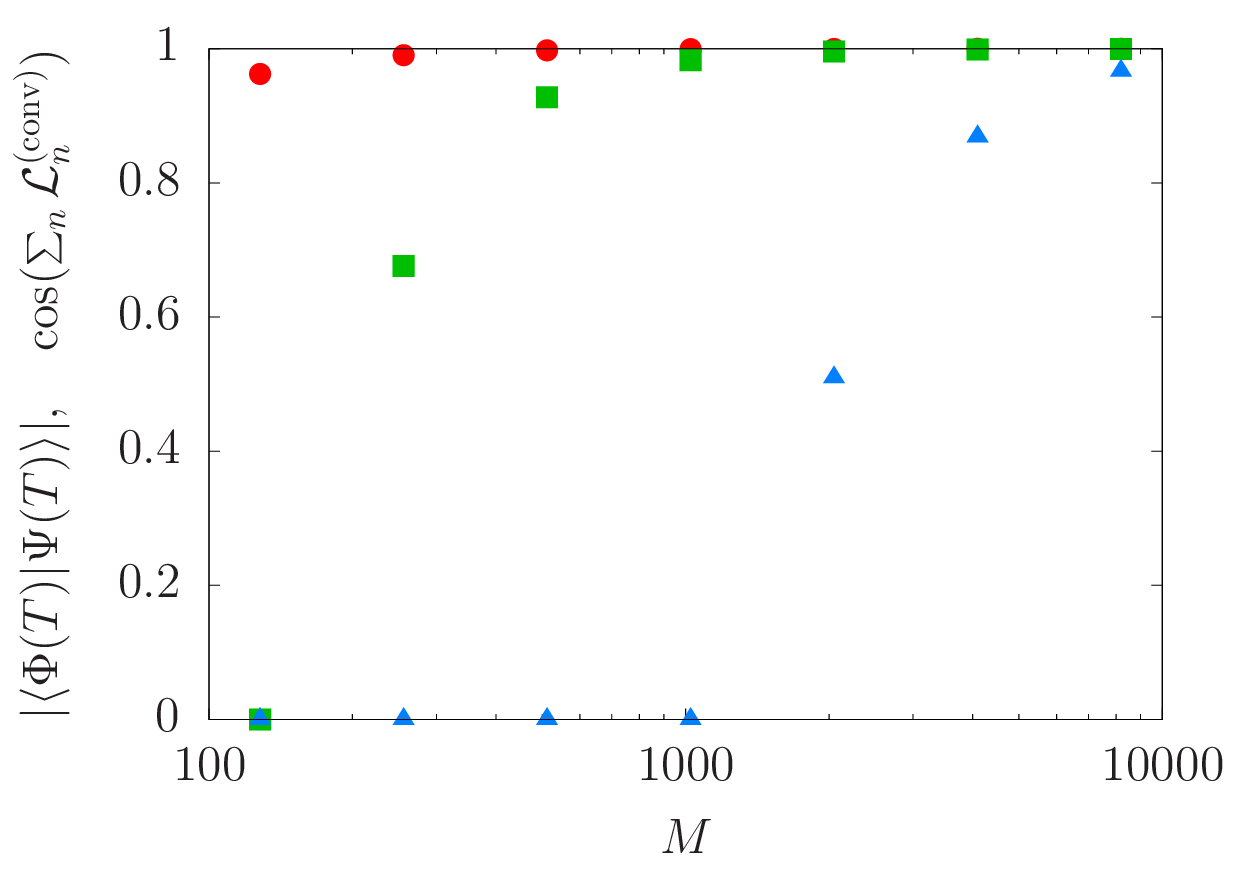}
\caption{\label{Fig.bd2}Overlap between the digital simulated dynamics and the true dynamics with its lower bounds against the number of the time slices $M$. Each symbol represents (red circles) the overlap $|\langle\Phi(T)|\Psi(T)\rangle|$, (green squares) our bound $\cos(\sum_n\mathcal{L}_n)$, and (blue triangles) the conventional bound $\cos(\sum_n\mathcal{L}_n^{\mathrm{conv}})$, respectively. Here, $T=10$.}
\end{figure}

Next, we change the number of the time slices $M$. 
Here, we set $T=10$, for which as we can find in Fig.~\ref{Fig.bd} both our bound and the conventional bound are useless when we set $M=T^2$. 
In Fig.~\ref{Fig.bd2}, the overlap between the digital simulated dynamics and the true dynamics for various number of the time slices $M$ is plotted with the lower bounds. 
Both bounds become tight for large number of the time slices, but our bound is significantly tighter than the conventional bound.

\textit{Error bound for noisy digital quantum simulation.---}
At last, we formally extend our formalism to error evaluation of noisy digital quantum simulation, i.e., we consider digital quantum simulation of the target state $\hat{\rho}(T)=|\Psi(T)\rangle\langle\Psi(T)|$ on noisy quantum computers. 
By introducing a unitary time-evolution map $\mathcal{D}_m[\cdot]=\hat{\mathcal{U}}(mT/M,(m-1)T/M)\cdot[\hat{\mathcal{U}}(mT/M,(m-1)T/M)]^\dag$ and a completely positive trace preserving map $\mathcal{E}[\cdot]$, which induces certain noise, we can express noisy digital simulated dynamics as
\begin{equation}
\begin{aligned}
\hat{\sigma}(nT/M)=\left(\prod_{m=n}^1\mathcal{E}\circ\mathcal{D}_m\right)[|0\rangle\langle0|],&\\
n=1,2,\dots,M,&
\end{aligned}
\end{equation}
for a small time interval $T/M$. 
In this case, distance between the noisy digital simulated dynamics $\hat{\sigma}(T)$ and the true dynamics $\hat{\rho}(T)$ is given by the Bures angle
\begin{equation}
\mathcal{L}(\hat{\sigma}(T),\hat{\rho}(T))=\arccos\left|\mathrm{Tr}\sqrt{\sqrt{\hat{\rho}(T)}\hat{\sigma}(T)\sqrt{\hat{\rho}(T)}}\right|, 
\end{equation}
which is the generalization of the Fubini-Study angle, i.e., the statistical distance between two mixed states~\cite{Braunstein1994}. 
By using the unitary invariance of the Bures angle, it can be rewritten as $\mathcal{L}(\hat{\sigma}(T),\hat{\rho}(T))=\mathcal{L}(\hat{\tau}(T),|0\rangle\langle0|)$, where $\hat{\tau}(nT/M)=[\hat{U}(nT/M,0)]^\dag\hat{\sigma}(nT/M)\hat{U}(nT/M,0)$. 
Now, as in the case of the Fubini-Study angle, we use the triangle inequality of the distance, and then we obtain an upper bound
\begin{equation}
\mathcal{L}(\hat{\sigma}(T),\hat{\rho}(T))\le\sum_{n=1}^M\mathcal{L}_n,
\end{equation}
where
\begin{equation}
\mathcal{L}_n=\mathcal{L}(\mathcal{F}[\hat{\sigma}((n-1)T/M)],\hat{\sigma}((n-1)T/M)),
\end{equation}
and
\begin{equation}
\begin{aligned}
\mathcal{F}[\hat{\sigma}((n-1)T/M)]=&[\hat{U}(nT/M,(n-1)T/M)]^\dag\\
&\times(\mathcal{E}\circ\mathcal{D}_n)[\hat{\sigma}((n-1)T/M)]\\
&\times\hat{U}(nT/M,(n-1)T/M).
\end{aligned}
\end{equation}
As in the case of the Fubini-Study angle, this upper bound is equivalent to a lower bound for overlap between the noisy digital simulated dynamics and the true dynamics.

As an example, we calculate each distance $\mathcal{L}_n$ by using the first-order Suzuki-Trotter decomposition and by adopting depolarizing noise
\begin{equation}
\mathcal{E}[\cdot]=(1-p)\cdot+p\frac{\hat{\mathbb{1}}}{D},
\end{equation}
where $D$ is the dimension of the present Hilbert space and $p$ represents the ratio of depolarization. 
Since this noise channel is induced for each small time interval $T/M$, we assume that $p=\gamma T/\hbar M$, where $\gamma$ is the decay rate. 
Then, each distance $\mathcal{L}_n$ is given by
\begin{equation}
\begin{aligned}
\mathcal{L}_n\approx&\Bigg[\frac{1}{2}\left(\frac{T^2}{2\hbar^2M^2}\right)^2\sum_{\substack{i,j \\ (p_i+p_j\neq0)}}\frac{(p_i-p_j)^2}{p_i+p_j}|\langle i|\hat{A}(nT/M)|j\rangle|^2\\
&+\frac{1}{2}\left(\frac{\gamma T}{\hbar M}\right)^2\sum_{\substack{i \\ (p_i\ne0)}}\frac{(p_i-1/D)^2}{2p_i}\Bigg]^{1/2},
\end{aligned}
\end{equation}
where $p_i$ and $|i\rangle$ is given by the spectral decomposition of the noisy digital simulated dynamics
\begin{equation}
\hat{\sigma}((n-1)T/M)=\sum_ip_i|i\rangle\langle i|. 
\end{equation}
It is also possible to further improve the first term as well as the noiseless case.

\textit{Summary.---}
In this paper, we introduced distance between the true dynamics and its digital simulated dynamics at the final time. 
Then, we derived its upper bound which is equivalent to the lower bound for the overlap between these two dynamics. 
We showed both in the analytical way and the numerical way that our result gives the more straightforward and tighter bound for error evaluation of digital quantum simulation than conventional approaches. 
We also extended our formalism to error evaluation of noisy digital quantum simulation, i.e., digital quantum simulation on noisy quantum computers. 
In the noisy case, spectral information of the noisy digital simulated dynamics is required for calculating the bound, but it is generally a hard task. 
Therefore, it is important future work to develop a way for efficiently calculating this bound. 
Methods of efficient quantum state tomography~\cite{Cramer2010} may resolve this problem.

\bibliography{Trotterbib}

\begin{thebibliography}{20}%
\makeatletter
\providecommand \@ifxundefined [1]{%
 \@ifx{#1\undefined}
}%
\providecommand \@ifnum [1]{%
 \ifnum #1\expandafter \@firstoftwo
 \else \expandafter \@secondoftwo
 \fi
}%
\providecommand \@ifx [1]{%
 \ifx #1\expandafter \@firstoftwo
 \else \expandafter \@secondoftwo
 \fi
}%
\providecommand \natexlab [1]{#1}%
\providecommand \enquote  [1]{``#1''}%
\providecommand \bibnamefont  [1]{#1}%
\providecommand \bibfnamefont [1]{#1}%
\providecommand \citenamefont [1]{#1}%
\providecommand \href@noop [0]{\@secondoftwo}%
\providecommand \href [0]{\begingroup \@sanitize@url \@href}%
\providecommand \@href[1]{\@@startlink{#1}\@@href}%
\providecommand \@@href[1]{\endgroup#1\@@endlink}%
\providecommand \@sanitize@url [0]{\catcode `\\12\catcode `\$12\catcode
  `\&12\catcode `\#12\catcode `\^12\catcode `\_12\catcode `\%12\relax}%
\providecommand \@@startlink[1]{}%
\providecommand \@@endlink[0]{}%
\providecommand \url  [0]{\begingroup\@sanitize@url \@url }%
\providecommand \@url [1]{\endgroup\@href {#1}{\urlprefix }}%
\providecommand \urlprefix  [0]{URL }%
\providecommand \Eprint [0]{\href }%
\providecommand \doibase [0]{https://doi.org/}%
\providecommand \selectlanguage [0]{\@gobble}%
\providecommand \bibinfo  [0]{\@secondoftwo}%
\providecommand \bibfield  [0]{\@secondoftwo}%
\providecommand \translation [1]{[#1]}%
\providecommand \BibitemOpen [0]{}%
\providecommand \bibitemStop [0]{}%
\providecommand \bibitemNoStop [0]{.\EOS\space}%
\providecommand \EOS [0]{\spacefactor3000\relax}%
\providecommand \BibitemShut  [1]{\csname bibitem#1\endcsname}%
\let\auto@bib@innerbib\@empty
\bibitem [{\citenamefont {Feynman}(1982)}]{Feynman1982}%
  \BibitemOpen
  \bibfield  {author} {\bibinfo {author} {\bibfnamefont {R.~P.}\ \bibnamefont
  {Feynman}},\ }\bibfield  {title} {\bibinfo {title} {{Simulating physics with
  computers}},\ }\href {https://doi.org/10.1007/BF02650179} {\bibfield
  {journal} {\bibinfo  {journal} {International Journal of Theoretical
  Physics}\ }\textbf {\bibinfo {volume} {21}},\ \bibinfo {pages} {467}
  (\bibinfo {year} {1982})}\BibitemShut {NoStop}%
\bibitem [{\citenamefont {Lloyd}(1996)}]{Lloyd1996}%
  \BibitemOpen
  \bibfield  {author} {\bibinfo {author} {\bibfnamefont {S.}~\bibnamefont
  {Lloyd}},\ }\bibfield  {title} {\bibinfo {title} {{Universal Quantum
  Simulators}},\ }\href {https://doi.org/10.1126/SCIENCE.273.5278.1073}
  {\bibfield  {journal} {\bibinfo  {journal} {Science}\ }\textbf {\bibinfo
  {volume} {273}},\ \bibinfo {pages} {1073} (\bibinfo {year}
  {1996})}\BibitemShut {NoStop}%
\bibitem [{\citenamefont {Wiebe}\ \emph {et~al.}(2010)\citenamefont {Wiebe},
  \citenamefont {Berry}, \citenamefont {H{\o}yer},\ and\ \citenamefont
  {Sanders}}]{Wiebe2010}%
  \BibitemOpen
  \bibfield  {author} {\bibinfo {author} {\bibfnamefont {N.}~\bibnamefont
  {Wiebe}}, \bibinfo {author} {\bibfnamefont {D.}~\bibnamefont {Berry}},
  \bibinfo {author} {\bibfnamefont {P.}~\bibnamefont {H{\o}yer}},\ and\
  \bibinfo {author} {\bibfnamefont {B.~C.}\ \bibnamefont {Sanders}},\
  }\bibfield  {title} {\bibinfo {title} {{Higher order decompositions of
  ordered operator exponentials}},\ }\href
  {https://doi.org/10.1088/1751-8113/43/6/065203} {\bibfield  {journal}
  {\bibinfo  {journal} {Journal of Physics A: Mathematical and Theoretical}\
  }\textbf {\bibinfo {volume} {43}},\ \bibinfo {pages} {065203} (\bibinfo
  {year} {2010})}\BibitemShut {NoStop}%
\bibitem [{\citenamefont {Poulin}\ \emph {et~al.}(2011)\citenamefont {Poulin},
  \citenamefont {Qarry}, \citenamefont {Somma},\ and\ \citenamefont
  {Verstraete}}]{Poulin2011}%
  \BibitemOpen
  \bibfield  {author} {\bibinfo {author} {\bibfnamefont {D.}~\bibnamefont
  {Poulin}}, \bibinfo {author} {\bibfnamefont {A.}~\bibnamefont {Qarry}},
  \bibinfo {author} {\bibfnamefont {R.}~\bibnamefont {Somma}},\ and\ \bibinfo
  {author} {\bibfnamefont {F.}~\bibnamefont {Verstraete}},\ }\bibfield  {title}
  {\bibinfo {title} {{Quantum Simulation of Time-Dependent Hamiltonians and the
  Convenient Illusion of Hilbert Space}},\ }\href
  {https://doi.org/10.1103/PhysRevLett.106.170501} {\bibfield  {journal}
  {\bibinfo  {journal} {Physical Review Letters}\ }\textbf {\bibinfo {volume}
  {106}},\ \bibinfo {pages} {170501} (\bibinfo {year} {2011})}\BibitemShut
  {NoStop}%
\bibitem [{\citenamefont {Somma}(2016)}]{Somma2016}%
  \BibitemOpen
  \bibfield  {author} {\bibinfo {author} {\bibfnamefont {R.~D.}\ \bibnamefont
  {Somma}},\ }\bibfield  {title} {\bibinfo {title} {{A Trotter-Suzuki
  approximation for Lie groups with applications to Hamiltonian simulation}},\
  }\href {https://doi.org/10.1063/1.4952761} {\bibfield  {journal} {\bibinfo
  {journal} {Journal of Mathematical Physics}\ }\textbf {\bibinfo {volume}
  {57}},\ \bibinfo {pages} {062202} (\bibinfo {year} {2016})}\BibitemShut
  {NoStop}%
\bibitem [{\citenamefont {Hadfield}\ and\ \citenamefont
  {Papageorgiou}(2018)}]{Hadfield2018}%
  \BibitemOpen
  \bibfield  {author} {\bibinfo {author} {\bibfnamefont {S.}~\bibnamefont
  {Hadfield}}\ and\ \bibinfo {author} {\bibfnamefont {A.}~\bibnamefont
  {Papageorgiou}},\ }\bibfield  {title} {\bibinfo {title} {{Divide and conquer
  approach to quantum Hamiltonian simulation}},\ }\href
  {https://doi.org/10.1088/1367-2630/aab1ef} {\bibfield  {journal} {\bibinfo
  {journal} {New Journal of Physics}\ }\textbf {\bibinfo {volume} {20}},\
  \bibinfo {pages} {043003} (\bibinfo {year} {2018})}\BibitemShut {NoStop}%
\bibitem [{\citenamefont {Childs}\ \emph {et~al.}(2019)\citenamefont {Childs},
  \citenamefont {Ostrander},\ and\ \citenamefont {Su}}]{Childs2019}%
  \BibitemOpen
  \bibfield  {author} {\bibinfo {author} {\bibfnamefont {A.~M.}\ \bibnamefont
  {Childs}}, \bibinfo {author} {\bibfnamefont {A.}~\bibnamefont {Ostrander}},\
  and\ \bibinfo {author} {\bibfnamefont {Y.}~\bibnamefont {Su}},\ }\bibfield
  {title} {\bibinfo {title} {{Faster quantum simulation by randomization}},\
  }\href {https://doi.org/10.22331/q-2019-09-02-182} {\bibfield  {journal}
  {\bibinfo  {journal} {Quantum}\ }\textbf {\bibinfo {volume} {3}},\ \bibinfo
  {pages} {182} (\bibinfo {year} {2019})}\BibitemShut {NoStop}%
\bibitem [{\citenamefont {Childs}\ \emph {et~al.}(2021)\citenamefont {Childs},
  \citenamefont {Su}, \citenamefont {Tran}, \citenamefont {Wiebe},\ and\
  \citenamefont {Zhu}}]{Childs2021}%
  \BibitemOpen
  \bibfield  {author} {\bibinfo {author} {\bibfnamefont {A.~M.}\ \bibnamefont
  {Childs}}, \bibinfo {author} {\bibfnamefont {Y.}~\bibnamefont {Su}}, \bibinfo
  {author} {\bibfnamefont {M.~C.}\ \bibnamefont {Tran}}, \bibinfo {author}
  {\bibfnamefont {N.}~\bibnamefont {Wiebe}},\ and\ \bibinfo {author}
  {\bibfnamefont {S.}~\bibnamefont {Zhu}},\ }\bibfield  {title} {\bibinfo
  {title} {{Theory of Trotter Error with Commutator Scaling}},\ }\href
  {https://doi.org/10.1103/PhysRevX.11.011020} {\bibfield  {journal} {\bibinfo
  {journal} {Physical Review X}\ }\textbf {\bibinfo {volume} {11}},\ \bibinfo
  {pages} {011020} (\bibinfo {year} {2021})}\BibitemShut {NoStop}%
\bibitem [{\citenamefont {Suzuki}(1976)}]{Suzuki1976}%
  \BibitemOpen
  \bibfield  {author} {\bibinfo {author} {\bibfnamefont {M.}~\bibnamefont
  {Suzuki}},\ }\bibfield  {title} {\bibinfo {title} {{Generalized Trotter's
  formula and systematic approximants of exponential operators and inner
  derivations with applications to many-body problems}},\ }\href
  {https://doi.org/10.1007/BF01609348} {\bibfield  {journal} {\bibinfo
  {journal} {Communications in Mathematical Physics}\ }\textbf {\bibinfo
  {volume} {51}},\ \bibinfo {pages} {183} (\bibinfo {year} {1976})}\BibitemShut
  {NoStop}%
\bibitem [{\citenamefont {Suzuki}(1985)}]{Suzuki1985}%
  \BibitemOpen
  \bibfield  {author} {\bibinfo {author} {\bibfnamefont {M.}~\bibnamefont
  {Suzuki}},\ }\bibfield  {title} {\bibinfo {title} {{Decomposition formulas of
  exponential operators and Lie exponentials with some applications to quantum
  mechanics and statistical physics}},\ }\href
  {https://doi.org/10.1063/1.526596} {\bibfield  {journal} {\bibinfo  {journal}
  {Journal of Mathematical Physics}\ }\textbf {\bibinfo {volume} {26}},\
  \bibinfo {pages} {601} (\bibinfo {year} {1985})}\BibitemShut {NoStop}%
\bibitem [{\citenamefont {Suzuki}(1991)}]{Suzuki1991}%
  \BibitemOpen
  \bibfield  {author} {\bibinfo {author} {\bibfnamefont {M.}~\bibnamefont
  {Suzuki}},\ }\bibfield  {title} {\bibinfo {title} {{General theory of fractal
  path integrals with applications to many-body theories and statistical
  physics}},\ }\href {https://doi.org/10.1063/1.529425} {\bibfield  {journal}
  {\bibinfo  {journal} {Journal of Mathematical Physics}\ }\textbf {\bibinfo
  {volume} {32}},\ \bibinfo {pages} {400} (\bibinfo {year} {1991})}\BibitemShut
  {NoStop}%
\bibitem [{\citenamefont {Huyghebaert}\ and\ \citenamefont
  {Raedt}(1990)}]{Huyghebaert1990}%
  \BibitemOpen
  \bibfield  {author} {\bibinfo {author} {\bibfnamefont {J.}~\bibnamefont
  {Huyghebaert}}\ and\ \bibinfo {author} {\bibfnamefont {H.~D.}\ \bibnamefont
  {Raedt}},\ }\bibfield  {title} {\bibinfo {title} {{Product formula methods
  for time-dependent Schrodinger problems}},\ }\href
  {https://doi.org/10.1088/0305-4470/23/24/019} {\bibfield  {journal} {\bibinfo
   {journal} {Journal of Physics A: Mathematical and General}\ }\textbf
  {\bibinfo {volume} {23}},\ \bibinfo {pages} {5777} (\bibinfo {year}
  {1990})}\BibitemShut {NoStop}%
\bibitem [{\citenamefont {Suzuki}\ and\ \citenamefont
  {Takahashi}(2020)}]{Suzuki2020}%
  \BibitemOpen
  \bibfield  {author} {\bibinfo {author} {\bibfnamefont {K.}~\bibnamefont
  {Suzuki}}\ and\ \bibinfo {author} {\bibfnamefont {K.}~\bibnamefont
  {Takahashi}},\ }\bibfield  {title} {\bibinfo {title} {{Performance evaluation
  of adiabatic quantum computation via quantum speed limits and possible
  applications to many-body systems}},\ }\href
  {https://doi.org/10.1103/PhysRevResearch.2.032016} {\bibfield  {journal}
  {\bibinfo  {journal} {Physical Review Research}\ }\textbf {\bibinfo {volume}
  {2}},\ \bibinfo {pages} {032016(R)} (\bibinfo {year} {2020})}\BibitemShut
  {NoStop}%
\bibitem [{\citenamefont {Hatomura}\ and\ \citenamefont
  {Takahashi}(2021)}]{Hatomura2021}%
  \BibitemOpen
  \bibfield  {author} {\bibinfo {author} {\bibfnamefont {T.}~\bibnamefont
  {Hatomura}}\ and\ \bibinfo {author} {\bibfnamefont {K.}~\bibnamefont
  {Takahashi}},\ }\bibfield  {title} {\bibinfo {title} {{Controlling and
  exploring quantum systems by algebraic expression of adiabatic gauge
  potential}},\ }\href {https://doi.org/10.1103/PhysRevA.103.012220} {\bibfield
   {journal} {\bibinfo  {journal} {Physical Review A}\ }\textbf {\bibinfo
  {volume} {103}},\ \bibinfo {pages} {012220} (\bibinfo {year}
  {2021})}\BibitemShut {NoStop}%
\bibitem [{\citenamefont {Funo}\ \emph {et~al.}(2021)\citenamefont {Funo},
  \citenamefont {Lambert},\ and\ \citenamefont {Nori}}]{Funo2021}%
  \BibitemOpen
  \bibfield  {author} {\bibinfo {author} {\bibfnamefont {K.}~\bibnamefont
  {Funo}}, \bibinfo {author} {\bibfnamefont {N.}~\bibnamefont {Lambert}},\ and\
  \bibinfo {author} {\bibfnamefont {F.}~\bibnamefont {Nori}},\ }\bibfield
  {title} {\bibinfo {title} {{General Bound on the Performance of
  Counter-Diabatic Driving Acting on Dissipative Spin Systems}},\ }\href
  {https://doi.org/10.1103/PhysRevLett.127.150401} {\bibfield  {journal}
  {\bibinfo  {journal} {Physical Review Letters}\ }\textbf {\bibinfo {volume}
  {127}},\ \bibinfo {pages} {150401} (\bibinfo {year} {2021})}\BibitemShut
  {NoStop}%
\bibitem [{\citenamefont {Hatomura}(2021)}]{Hatomura2021a}%
  \BibitemOpen
  \bibfield  {author} {\bibinfo {author} {\bibfnamefont {T.}~\bibnamefont
  {Hatomura}},\ }\bibfield  {title} {\bibinfo {title} {{Performance evaluation
  of invariant-based inverse engineering by quantum speed limit}},\ }\href
  {http://arxiv.org/abs/2112.07253} {\  (\bibinfo {year} {2021})},\ \Eprint
  {https://arxiv.org/abs/2112.07253} {arXiv:2112.07253} \BibitemShut {NoStop}%
\bibitem [{\citenamefont {Takahashi}(2021)}]{Takahashi2021}%
  \BibitemOpen
  \bibfield  {author} {\bibinfo {author} {\bibfnamefont {K.}~\bibnamefont
  {Takahashi}},\ }\bibfield  {title} {\bibinfo {title} {{Quantum lower and
  upper speed limits}},\ }\href {http://arxiv.org/abs/2112.12631} {\  (\bibinfo
  {year} {2021})},\ \Eprint {https://arxiv.org/abs/2112.12631}
  {arXiv:2112.12631} \BibitemShut {NoStop}%
\bibitem [{\citenamefont {Wootters}(1981)}]{Wootters1981}%
  \BibitemOpen
  \bibfield  {author} {\bibinfo {author} {\bibfnamefont {W.~K.}\ \bibnamefont
  {Wootters}},\ }\bibfield  {title} {\bibinfo {title} {{Statistical distance
  and Hilbert space}},\ }\href {https://doi.org/10.1103/PhysRevD.23.357}
  {\bibfield  {journal} {\bibinfo  {journal} {Physical Review D}\ }\textbf
  {\bibinfo {volume} {23}},\ \bibinfo {pages} {357} (\bibinfo {year}
  {1981})}\BibitemShut {NoStop}%
\bibitem [{\citenamefont {Braunstein}\ and\ \citenamefont
  {Caves}(1994)}]{Braunstein1994}%
  \BibitemOpen
  \bibfield  {author} {\bibinfo {author} {\bibfnamefont {S.~L.}\ \bibnamefont
  {Braunstein}}\ and\ \bibinfo {author} {\bibfnamefont {C.~M.}\ \bibnamefont
  {Caves}},\ }\bibfield  {title} {\bibinfo {title} {{Statistical distance and
  the geometry of quantum states}},\ }\href
  {https://doi.org/10.1103/PhysRevLett.72.3439} {\bibfield  {journal} {\bibinfo
   {journal} {Physical Review Letters}\ }\textbf {\bibinfo {volume} {72}},\
  \bibinfo {pages} {3439} (\bibinfo {year} {1994})}\BibitemShut {NoStop}%
\bibitem [{\citenamefont {Cramer}\ \emph {et~al.}(2010)\citenamefont {Cramer},
  \citenamefont {Plenio}, \citenamefont {Flammia}, \citenamefont {Somma},
  \citenamefont {Gross}, \citenamefont {Bartlett}, \citenamefont
  {Landon-Cardinal}, \citenamefont {Poulin},\ and\ \citenamefont
  {Liu}}]{Cramer2010}%
  \BibitemOpen
  \bibfield  {author} {\bibinfo {author} {\bibfnamefont {M.}~\bibnamefont
  {Cramer}}, \bibinfo {author} {\bibfnamefont {M.~B.}\ \bibnamefont {Plenio}},
  \bibinfo {author} {\bibfnamefont {S.~T.}\ \bibnamefont {Flammia}}, \bibinfo
  {author} {\bibfnamefont {R.}~\bibnamefont {Somma}}, \bibinfo {author}
  {\bibfnamefont {D.}~\bibnamefont {Gross}}, \bibinfo {author} {\bibfnamefont
  {S.~D.}\ \bibnamefont {Bartlett}}, \bibinfo {author} {\bibfnamefont
  {O.}~\bibnamefont {Landon-Cardinal}}, \bibinfo {author} {\bibfnamefont
  {D.}~\bibnamefont {Poulin}},\ and\ \bibinfo {author} {\bibfnamefont {Y.-K.}\
  \bibnamefont {Liu}},\ }\bibfield  {title} {\bibinfo {title} {{Efficient
  quantum state tomography}},\ }\href {https://doi.org/10.1038/ncomms1147}
  {\bibfield  {journal} {\bibinfo  {journal} {Nature Communications}\ }\textbf
  {\bibinfo {volume} {1}},\ \bibinfo {pages} {149} (\bibinfo {year}
  {2010})}\BibitemShut {NoStop}%
\end{thebibliography}%

\end{document}